\documentclass[a4paper, twocolumn,superscriptaddress,showpacs]{revtex4}

\usepackage{amssymb}
\usepackage{amsmath}
\usepackage{epsfig}
\usepackage{color}
\usepackage{graphics, graphicx}
\usepackage{bbold}
\usepackage{psfrag}
\usepackage{mathcomp}
\usepackage{subfigure}
\usepackage{verbatim}
\usepackage{color}
\usepackage[colorlinks,citecolor=blue]{hyperref}

\begin{document}

\date{\today}
\title{Bose-Einstein condensate in an optical lattice with Raman-assisted two-dimensional spin-orbit coupling}
\author{Jian-Song Pan}
\affiliation{Key Laboratory of Quantum Information, University of Science and Technology of China, CAS, Hefei, Anhui, 230026, China}
\affiliation{Synergetic Innovation Center of Quantum Information and Quantum Physics, University of Science and Technology of China, Hefei, Anhui 230026, China}
\author{Wei Zhang}
\email{wzhangl@ruc.edu.cn}
\affiliation{Department of Physics, Renmin University of China, Beijing 100872, China}
\affiliation{Beijing Key Laboratory of Opto-electronic Functional Materials and Micro-nano Devices,
Renmin University of China, Beijing 100872, China}
\author{Wei Yi}
\email{wyiz@ustc.edu.cn}
\affiliation{Key Laboratory of Quantum Information, University of Science and Technology of China, CAS, Hefei, Anhui, 230026, China}
\affiliation{Synergetic Innovation Center of Quantum Information and Quantum Physics, University of Science and Technology of China, Hefei, Anhui 230026, China}
\author{Guang-Can Guo}
\affiliation{Key Laboratory of Quantum Information, University of Science and Technology of China, CAS, Hefei, Anhui, 230026, China}
\affiliation{Synergetic Innovation Center of Quantum Information and Quantum Physics, University of Science and Technology of China, Hefei, Anhui 230026, China}

\begin{abstract}
In a recent experiment by Wu {\textit et al.} (arXiv:1511.08170), a Raman-assisted two-dimensional spin-orbit coupling has been realized for a Bose-Einstein condensate in an optical lattice potential. In light of this exciting progress, we study in detail key properties of the system. As the Raman lasers inevitably couple atoms to high-lying bands, the behaviors of the system in both the single- and many-particle sectors are significantly affected. In particular, the high-band effects enhance the plane-wave phase and lead to the emergence of ``roton'' gaps at low Zeeman fields. Furthermore, we identify high-band-induced topological phase boundaries in both the single-particle and the quasi-particle spectra. We then derive an effective two-band model, which captures the high-band physics in the experimentally relevant regime. Our results not only offer valuable insights into the novel two-dimensional lattice spin-orbit coupling, but also provide a systematic formalism to model high-band effects in lattice systems with Raman-assisted spin-orbit couplings.
\end{abstract}
%whose parameters can be systematically determined from full-band calculations.

\maketitle

%%%%%%%%%%%%%%
%%%%%%%%%%%%%%
\emph{Introduction}.--
Synthetic spin-orbit coupling (SOC) in cold atomic gases has stimulated much research interest ever since its experimental implementation~\cite{gauge2exp,fermisocexp1,fermisocexp2,socreview1,socreview2,socreview3,socreview4,socreview5,socreview6,socreview7}. As SOC plays a key role in solid-state materials such as topological insulators and topological superconductors~\cite{kanereview,zhangscreview}, the realization of synthetic SOC in cold atoms offers a brand new platform for the investigation of these exotic states.
In a very recent experiment, a Raman-assisted two-dimensional (2D) SOC has been realized for a Bose-Einstein condensate (BEC) loaded in an optical lattice potential~\cite{Wu2015,Liu2014}. Together with the experimental implementation of 2D SOC in harmonically trapped Fermi gases~\cite{2dsoczhang1,2dsoczhang2}, these achievements represent a major step toward the quantum emulation of topological matter in ultracold atoms. To move forward on solid grounds, a thorough understanding of the properties of cold atoms in these novel 2D SOC setups is essential.

In this work, we study in detail key properties of a BEC under the 2D lattice SOC in Ref.~\cite{Wu2015}. By characterizing the single-particle spectrum, band topology, many-body phase diagram, and quasi-particle excitations, we gain valuable insights of the system. Importantly, the Raman-assisted SOC couple atoms to high-lying bands, which have significant impact on the system~\cite{cuisoc,wycavity,cavitylong}.
In particular, the high-band effects stabilize the plane-wave (PW) phase at low Zeeman fields, which lead to the opening of ``roton'' gaps in the excitation spectrum that are absent in a simple two-band model~\cite{rotontheory,rotonsoc,rotonchicago}. We also identify a high-band induced topological phase boundary for the lowest band of the single-particle spectrum, which is carried over to the bands in the Bogoliubov excitation spectrum. We propose to capture the high-band effects using an effective two-band Hamiltonian, whose parameters can be systematically determined from full-band calculations regardless of the form of SOC. We show that all the essential high-band-induced properties can be reproduced in the experimentally relevant regime using our effective model. Our results can be tested in current BEC experiments, and provide important insights that would be valuable for future experimental efforts toward a topological Fermi gas using 2D lattice SOC.

\emph{Model Hamiltonian}.--
Experimentally, the 2D lattice SOC is realized by coupling two states in the ground-state hyperfine manifold by two different Raman processes that are spatially dependent~\cite{Wu2015}. The single-particle Hamiltonian can be written as
\begin{equation}
\hat{\mathcal{H}}=\int d {\bf r}\left(\begin{array}{cc}
\hat{\psi}_{\uparrow}^{\dagger} & \hat{\psi}_{\downarrow}^{\dagger}\end{array}\right)\hat{h}_{0}\left(\begin{array}{c}
\hat{\psi}_{\uparrow}\\
\hat{\psi}_{\downarrow}
\end{array}\right)
\label{eqn:H}
\end{equation}
with
\begin{equation}
 \hat{h}_{0}=\left(\begin{array}{cc}
\frac{\hbar^{2}\hat{{\bf k}}^{2}}{2m}+m_{z} & M\\
M^{\ast} & \frac{\hbar^{2}\left(\hat{{\bf k}}-k_{0}\boldsymbol{e}_{x}-k_{0}\boldsymbol{e}_{y}\right)^{2}}{2m}-m_{z}
\end{array}\right)+V,
\label{eqn:h0}
\end{equation}
where $\hat{\psi}_{\sigma}$ ($\sigma=\uparrow,\downarrow$) are the field operators for fermions with pseudo-spins $\sigma$, $m_z$ is the effective Zeeman field, $m$ is the atomic mass, ${\bf k}$ is the 2D linear momentum in continuous space, and $\boldsymbol{e}_{x,y}$ are respectively the unit vectors in the $x$- and the $y$-direction. To define the pseudo-spins in Eq.~(\ref{eqn:H}), the gauge transformation $\hat{\psi}_{\downarrow}\longrightarrow e^{-ik_{0}\left(x+y\right)}\hat{\psi}_{\downarrow}$ has been applied to the hyperfine states of the atoms~\cite{Wu2015}. The lattice potential $V=-V_{0}\left[\cos^{2}\left(k_{0}x\right)+\cos^{2}\left(k_{0}y\right)\right]$, with a lattice depth $V_0$ and a lattice constant $a=\pi/k_0$. Without loss of generality, we take $V_0=4E_r$ throughout the work, where $E_r=\hbar^2k_0^2/2m$ is used as the unit of energy. The spatially dependent Raman potential $M=-M_{0}\left[\sin\left(k_{0}x\right)e^{-ik_{0}x}-i\sin\left(k_{0}y\right)e^{-ik_{0}y}\right]$ is crucial to the 2D SOC, where $M_0$ is related to the laser parameters. The 2D SOC arises from the two Raman processes characterized by the off-diagonal terms in Eq.~(\ref{eqn:h0}). Finally, due to the spin degrees of freedom, the $s$-orbital of the lattice potential consists of two bands.

\begin{figure}[tbp]
\includegraphics[width=9cm]{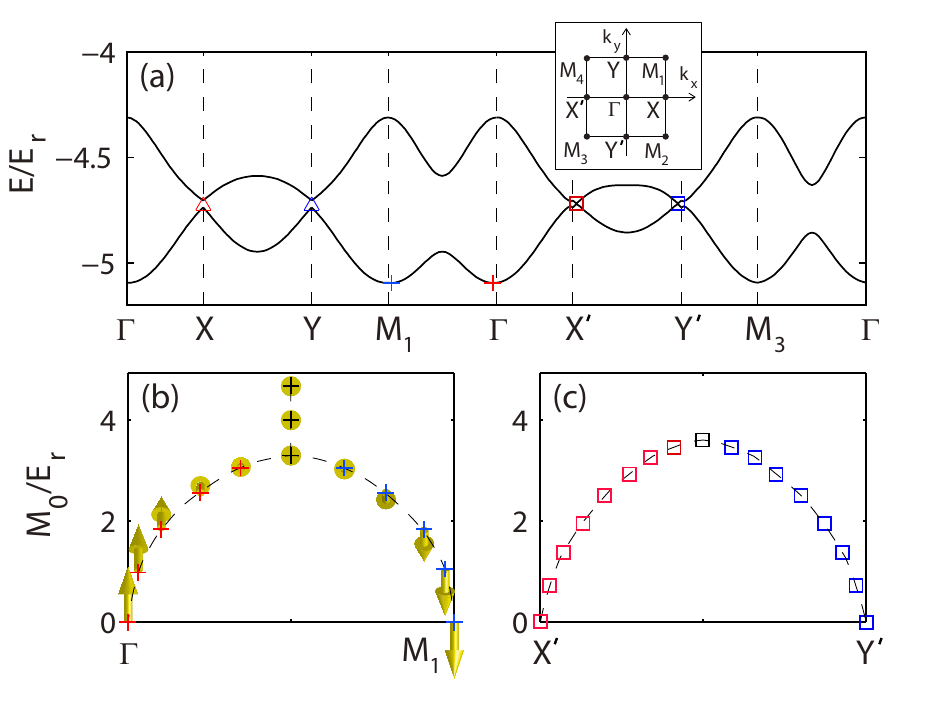}
\caption{(Color online) (a) The lowest two bands in the single-particle spectrum along a contour in the first Brillouin zone (1BZ) of the lattice potential $V$. At $M_0/E_r=1$, the Dirac points near X and Y are gapped out (triangle); while the remaining two Dirac points (squares) and the lowest-energy points (crosses) are all shifted away from their original positions at $M_0=0$. (b) Evolution of the location and pseudo-spin polarization (arrows) of the lowest-energy states with increasing $M_0$. (c) Evolution of the Dirac points with increasing $M_0$. For all subplots, $m_z=0$.}
\label{fig:sps}
\end{figure}

\emph{Single-particle spectrum and band topology}.--
The single-particle spectrum of the system can be obtained by diagonalizing Hamiltonian (\ref{eqn:H}). Two outstanding features of the lowest two bands in the $s$-orbital are the degeneracy of the lowest-energy states and the existence of Dirac points, as illustrated by crosses and squares in Fig.~\ref{fig:sps}a. Interestingly, for both the lowest-energy states and the Dirac points, their numbers and locations in momentum space depend sensitively on the strength of the Raman potential $M_0$.

In Fig.~\ref{fig:sps}b, we show the evolution of locations for the lowest-energy states (crosses) as the coupling strength $M_0$ changes. When $m_z=0$, the locations of the two degenerate lowest-energy states (red and blue crosses) move toward each other along the $\Gamma$--${\rm M}_1$ line with increasing $M_0$, until they merge at a critical $M_0\approx3.28E_r$, where the degeneracy is lifted. During this process, the spin orientations of the two states rotate within the plane perpendicular to the $\Gamma$--${\rm M}_1$ line as indicated by arrows in Fig.~\ref{fig:sps}b. Here, the spin orientation is defined as $\langle \hat{\boldsymbol{\sigma}}\rangle=\sum_i\langle \hat{\sigma}_i\rangle \boldsymbol{e}_{i}$ ($i=x,y,z$) with $\hat{\sigma}_i$ the Pauli matrices. When $m_z \neq 0$, the double degeneracy of the ground state would be lifted due to the explicitly broken of time-reversal symmetry. Following Eq.~(\ref{eqn:H}), for $m_z<0$, the lowest-energy state closer to the $\Gamma$ point (red cross in Fig.~\ref{fig:sps}a) would become the ground state.

While knowledge of the evolution of the lowest-energy points are important for the understanding of many-body phases and quasi-particle excitations of a BEC, the location and evolution of the Dirac points between the lowest two bands are crucial to the band topology. This would be particularly important when a Fermi gas is loaded into the lattice. At $m_z=0,M_0=0$, four Dirac points appear at ${\rm X}$, ${\rm Y}$, ${\rm X}'$ and ${\rm Y}'$ points. When $M_0$ becomes finite, the Dirac points at ${\rm X}$ and ${\rm Y}$ would be gapped out (see Fig.~\ref{fig:sps}a), and the remaining two would move toward each other along the ${\rm X}'$--${\rm Y}'$ line with increasing $M_0$ (red and blue squares in Fig.~\ref{fig:sps}c). These two Dirac points merge at $M_0\approx3.60E_r$, beyond which all Dirac points disappear. When $m_z$ becomes finite on the other hand, all the Dirac points would be gapped out as well.

The aforementioned features of the single-particle bands are crucially related to the high-band effects induced by SOC. To see this, we consider as a comparison a simple two-band model $\hat{h}_{00}=\boldsymbol{d}\cdot\hat{\boldsymbol{\sigma}}$ by completely neglecting the higher bands beyond the $s$-orbital. Here, $d_{x,y}=\pm2t_{\rm so}\sin\left(k_{y,x}a\right)$ and $d_{z}=m_{z}-2t_{s}\cos\left(k_{x}a\right)-2t_{s}\cos\left(k_{y}a\right)$
with quasi-momentum $\boldsymbol{k}$ and hopping coefficients $\{t_{s},t_{\rm so}\}$~\cite{suppl}.
The momentum-dependent hopping terms in $\boldsymbol{d}$ can be viewed as effective Zeeman fields, which is eventually responsible for the band topology. Under $\hat{h}_{00}$, we find that the locations of neither the lowest-energy states, nor the Dirac points, change as $M_0$ becomes finite~\cite{footnotenew}.
This finding is in stark contrast to the full-band calculations using Eq.~(\ref{eqn:H}), and is the direct result of Raman couplings with higher bands. In fact, as we will show later, such high-band effects induce a constant in-plane Zeeman field and explicitly break the inversion symmetry in the single-particle spectrum~\cite{suppl}.
As the single-particle ground state moves away from the $\Gamma$ point, its spin polarization evolves with increasing $M_0$ (see Fig.~\ref{fig:sps}b) under the effective Zeeman fields.

\emph{Ground-state phase diagram}.--
With a clear understanding of the single-particle spectrum, we now study the many-body ground-state phase diagram in the presence of intra- and inter-species interactions. The interaction Hamiltonian $\hat{V}_{\rm int}=\frac{1}{2}\sum_{\sigma\sigma^\prime}g_{\sigma\sigma^\prime}\int d {\bf r}\hat{\psi}_{\sigma}^{\dagger}\hat{\psi}_{\sigma^\prime}^{\dagger}\hat{\psi}_{\sigma^\prime}\hat{\psi}_{\sigma}$, where $g_{\sigma\sigma^\prime}$ is the interaction strength between spins $\sigma$ and $\sigma^\prime$. For convenience, we restrict our attention to the case of $g_{\uparrow\uparrow}=g_{\downarrow\downarrow}=g$ and $g_{\uparrow\downarrow}=g_{\downarrow\uparrow}\leq g$. In general, $g_{\uparrow\downarrow}\neq g$, and we may define a dimensionless parameter $\gamma=(g-g_{\uparrow\downarrow})/g$.

\begin{figure}[tbp]
\includegraphics[width=9cm]{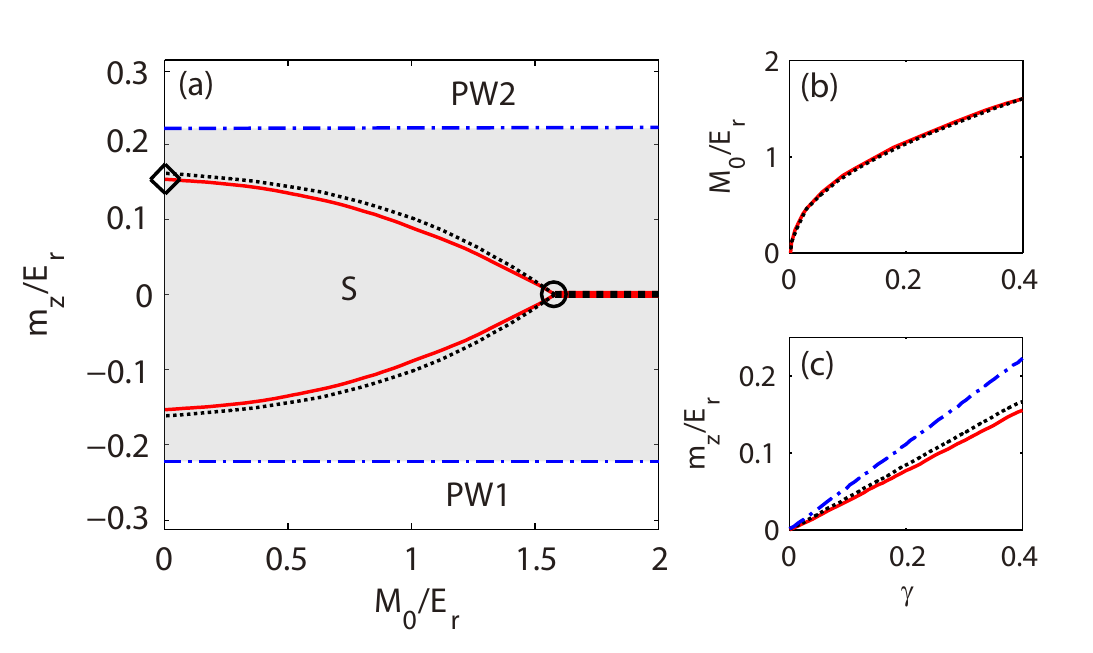}
\caption{(Color online) (a) Ground-state phase diagram at zero temperature for the stripe (S) phase, and the plane-wave (PW1 and PW2) phases. In the PW1 (PW2) phase, the atoms Bose condense near the lowest-energy single-particle state corresponding to the red (blue) crosses in Fig.~\ref{fig:sps}a. The red curves are from the full-band calculation, the blue dash-dotted curves are from a simple two-band model, and the black dotted curves are from the effective two-band model discussed later. The shaded region corresponds to the S phase from a simple two-band calculation. Note that the discrepancy between the dash-dotted and the solid curves at $M_0=0$ comes from nearest-neighbor interactions, which are absent from a simple two-band model here. (b)(c) Evolution of the critical points marked by circle (b), and  diamond (c), respectively. The line-shape conventions are the same as in (a). The interaction parameters are set as $\gamma=0.4$ and $gn_0/E_r=0.5$, where $n_0$ is the average number density.}
\label{fig:pd}
\end{figure}

We derive the Gross-Pitaevskii (GP) equation for the ground-state wave function of the BEC by minimizing the energy functional of the system~\cite{BECbook}
\begin{equation}
i\hbar\frac{\partial}{\partial t}\left(\begin{array}{c}
\phi_{\uparrow}\\
\phi_{\downarrow}
\end{array}\right)=\left(\hat{h}_{0}+h_{\rm int}\right)\left(\begin{array}{c}
\phi_{\uparrow}\\
\phi_{\downarrow}
\end{array}\right),
\label{eqn:GP}
\end{equation}
where
\begin{equation}
h_{\rm int}=\left(\begin{array}{cc}
g\left|\phi_{\uparrow}\right|^{2}+g_{\uparrow\downarrow}\left|\phi_{\downarrow}\right|^{2} & 0\\
0 & g_{\downarrow\uparrow}\left|\phi_{\uparrow}\right|^{2}+g\left|\phi_{\downarrow}\right|^{2}
\end{array}\right).
\label{eqn:hint}
\end{equation}
The ground-state wave function $\phi_{\sigma}$ can be calculated from Eq.~(\ref{eqn:GP}) using imaginary-time evolutions~\cite{imagtime}. In Fig.~\ref{fig:pd}, we show a typical ground-state phase diagram on the $M_0$--$m_z$ plane, where, driven by interactions, a two-fold degeneracy in the single-particle spectrum gives rise to the competition between two PW phases (PW1 and PW2) and a stripe (S) phase.
In contrast, under a simple two-band model $\hat{h}_{00}$ plus on-site interactions, the PW phase only emerges when the effective Zeeman field $m_z$ is large enough, which leads to a significantly over-estimated region for the S phase (shaded region in Fig.~\ref{fig:pd}a).
%In this sense, the couplings to higher bands effectively stabilize the PW phase.

\begin{figure}[tbp]
\includegraphics[width=9cm]{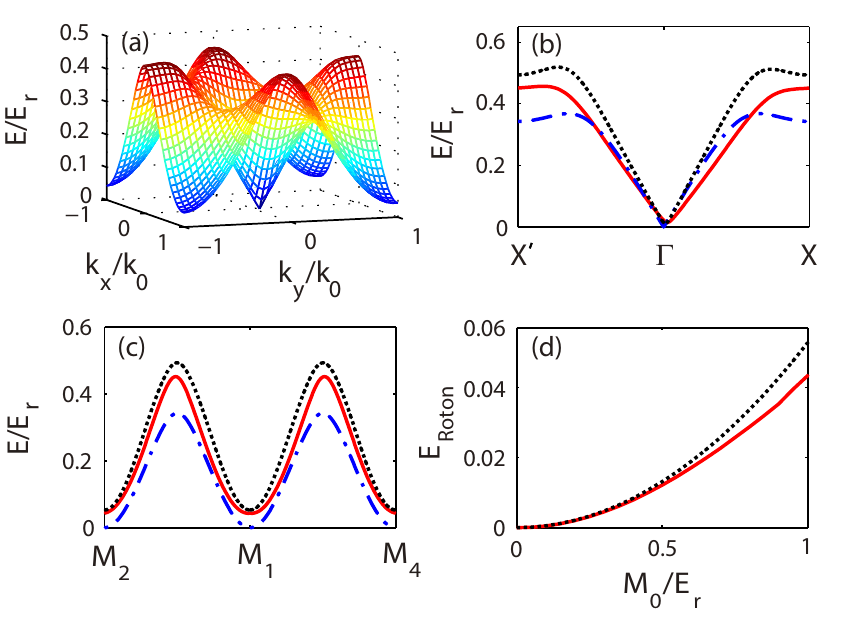}
\caption{(Color online) (a) The lowest band of the Bogoliubov spectrum from a full-band calculation~\cite{footnote}. (b)(c) The Bogoliubov spectrum along the ${\rm X}^\prime$--$\Gamma$--${\rm X}$ (b), and ${\rm M}_2$--${\rm M}_1$--${\rm M}_4$ (c) axes, respectively.  %Apparently, the ``roton'' gap at $M_1$ is absent in the simple two-band model (blue dash-dotted), while present in the full-band calculation (red solid)
(d) Variations of the ``roton'' gap at ${\rm M}_1$ as a function of $M_0$ with $m_z=0$. The parameters are $M_0/E_r=1$, $m_z=0$, $gn_0/E_r=0.2$ and $\gamma=0$. In (b)(c)(d), we show results under the full-band calculation (red solid), the simple two-band model (blue dash-dotted), and the effective two-band model discussed later in the text (black dotted). }
\label{fig:bs}
\end{figure}

\emph{Bogoliubov spectrum}.--
Based on the many-body ground-state phase diagram, we now investigate the Bogoliubov spectrum. For simplicity, we assume that the system is in a PW state, where the atoms Bose condense near a single-particle lowest-energy state. Following the standard Bogoliubov theory, we expand the field operator as $\hat{\psi}_{\sigma}=\phi_{\sigma} +\delta\hat{\psi}_{\sigma}$, with the fluctuation $\delta\hat{\psi}_{\sigma}$ above the ground state. We then calculate the quasi-particle spectrum by diagonalizing the Bogoliubov Hamiltonian for the fluctuations using paraunitary transformations~\cite{jpb}. In Fig.~\ref{fig:bs}, we show the lowest band of a typical quasi-particle spectrum with the parameters: $gn_0/E_r=0.2$, $m_z=0$ and $\gamma=0$. A linear dispersion emerges at low energies, which can be identified as the gapless Goldstone mode (Figs.~\ref{fig:bs}a and \ref{fig:bs}b). Under the typical experimental parameter $M_0/E_r=1$~\cite{Wu2015}, the deviation of the condensation position from the $\Gamma$ point already manifests itself in the excitation spectrum, such that the position of the gapless Goldstone mode is not at the $\Gamma$ point (Fig.~\ref{fig:bs}b). Notably, the sound velocity of the system becomes anisotropic, which should affect the collective modes of the system. On the other hand, the gaps appearing near ${\rm M}_{i=1,2,3,4}$ points are reminiscent of the ``roton'' gaps in a BEC under the effective 1D SOC~\cite{rotontheory,rotonsoc,rotonchicago,rotonwsu}. As a comparison, we emphasize that, in a simple two-band model, the ``roton'' gaps are not present under the same parameters (see Fig.~\ref{fig:bs}c), but only emerge at a much larger $m_z$, when the PW phase becomes stable. We also find that the ``roton'' gap at ${\rm M}_1$ decreases as the phase boundary to the S phase is approached, and it vanishes at the phase boundary between the PW and the S phases, as shown in Fig.~\ref{fig:bs}d. Similar ``roton'' softening can be observed for the gaps near the ${\rm M}_2$, ${\rm M}_3$, ${\rm M}_4$ points, but due to the broken inversion symmetry, they remain finite as the gap near ${\rm M}_1$ vanishes and the system enters the S phase. Thus, the emergence and softening of the ``roton'' gap at small Zeeman field offers the valuable opportunity to observe the high-band effects under the current experimental parameters with a relatively small $M_0$.

\begin{figure}[tbp]
\includegraphics[width=9cm]{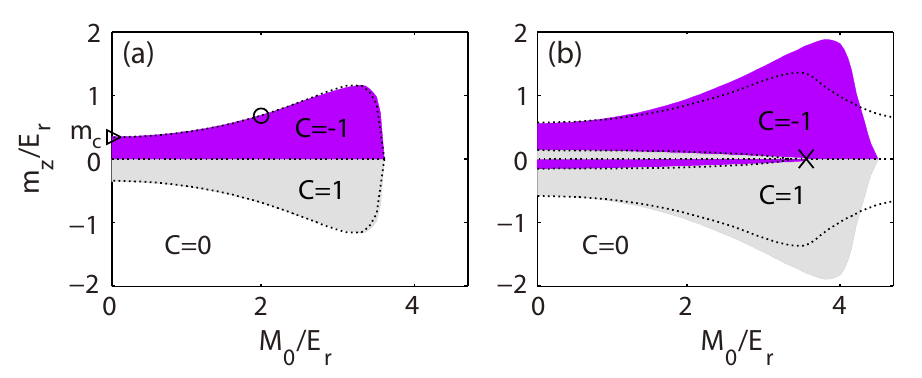}
\caption{(Color online) (a) The Chern number phase diagram of the lowest band in the single-particle spectrum on the $M_0$--$m_z$ plane. The triangle marks the topological boundary in a simple two-band model, which does not change as $M_0$ increases. (b) Chern numbers of the lowest band in the Bogoliubov spectrum. For (b), the parameters are the same as those in Fig.~\ref{fig:bs}. The fine phase structure near the $m_z=0$ axis ends at $M_0/E_r\approx 3.60$ (cross), where the Dirac points disappear. The black dotted curves are the corresponding boundaries calculated with an effective two-band model. The interaction parameters here are the same as those in Fig.~\ref{fig:bs}.}
\label{fig:topo}
\end{figure}

\emph{Topological band structure}.--
As demonstrated in Ref.~\cite{Wu2015}, the lowest band of Hamiltonian Eq.~(\ref{eqn:H}) can be topologically non-trivial under appropriate parameters. Under the two-band model $\hat{h}_{00}$, the gap between the lowest two bands closes at a critical Zeeman field $m_c=4t_{s}$, and the lowest band changes from topologically non-trivial for $|m_z|<m_c$ to topologically trivial for $|m_z|>m_c$~\cite{Liu2014,Wu2015}. Under the Raman-induced high-band effect however, the condition for the emergence of a topological band structure can be quite different. To characterize the topological invariant of the lowest band, we calculate the Chern number of the band~\cite{topocal1}:
$C=\frac{1}{2\pi i}\int_{T}d\boldsymbol{k}\left(\frac{\partial A_{y}}{\partial k_{x}}-\frac{\partial A_{x}}{\partial k_{y}}\right)$, where the Berry connection $A_{x,y}=\langle\phi_{0}\left(\boldsymbol{k}\right)|\frac{\partial}{\partial k_{x,y}}|\phi_{0}\left(\boldsymbol{k}\right)\rangle$ and $|\phi_0(\boldsymbol{k})\rangle$
is single-particle wave function at the quasi-momentum $\boldsymbol{k}$ in the first Brillioun zone. In Fig.~\ref{fig:topo}a, we plot the Chern-number phase diagram for the topology of the lowest band. Apparently, as $M_0$ increases, the boundaries of the topological band deviates from that of the simple two-band model. Interestingly, there appears to be a maximal $M_0$ below which the lowest band can be topological. For small $m_z$, this suggests the existence of a high-band induced topological boundary as $M_0$ is increased. These features can be probed using the spin-projection measurements~\cite{Wu2015}, and can give rise to new topological phase transitions in a Fermi gas under the 2D lattice SOC.

Another interesting observation is that the topology of the single-particle band structure can be carried over to the quasi-particle excitation spectrum. Unlike the single-particle situation, the Chern number for the excitation spectrum is defined in a symplectic manner~\cite{bogotopo1,bogotopo2}. In Fig.~\ref{fig:topo}b, we show the Chern-number diagram of the lowest band in the excitation spectrum. While the overall phase structure is similar to that of the lowest band in the single-particle spectrum, the topologically non-trivial region is larger in the excitation spectrum. Furthermore, as the spin polarization of the condensate differs on either side of the $m_z=0$ axis, a fine structure emerges in the phase boundaries near the $m_z=0$ axis at small $M_0$, which may be attributed to the competition between the interactions and the effective Zeeman field $m_z$~\cite{footnote2}. The non-trivial band topology of the excitation spectrum can lead to interesting chiral edge modes in the quasi-particle excitations in the presence of boundaries~\cite{bogotopo2}. In cold atomic experiment, these edge modes may be excited and probed using radio frequency spectroscopy.

%Comparing the topological phase boundary under the full-band calculation with those under the simple and the effective two-band models, we see that while the simple two-band model features a constant phase boundary in $M_0$, the effective two-band model reproduces the phase boundary for the experimentally relevant region $M_0<2E_r$. For larger $M_0$, the effective model deviates from the full-band model, suggesting the impact of higher-order corrections.

\emph{Effective two-band model}.--
So far, we have demonstrated that high-band effects can significantly modify the system properties as $M_0$ increases. However, it is not always easy or desirable to carry out a full-band calculation. To solve this dilemma, we propose an effective two-band model, which captures the essential high-band physics in the experimentally relevant regime when $M_0$ is not too large. Our formalism can be summarized as follows: (i) we
start from a minimal multi-band model with only $s$- and $p$-orbital of the lattice potential, and perform adiabatic elimination to get an effective two-band model with only $s$-orbital contributions; (ii) we fit the parameters of the effective two-band model using a handful of data points from the single-particle spectrum and the many-body phase diagram obtained by the full-band calculations. The consideration behind this procedure is that the effective model derived in (i), while applicable in the limit of a deep lattice and weak Raman couplings, should contain the leading terms responsible for the high-band effects, although the coefficients may be modified as $M_0$ increases. Formally, the effective model can be written as
\begin{eqnarray}
\label{eqn:Heff}
\hat{H}_{\rm eff}&=&\sum_{\boldsymbol{k}}\hat{\beta}^{\dag}_{\boldsymbol{k}}(\delta_{\rm eff}+{\boldsymbol{d}_{\rm eff}}\cdot\hat{\boldsymbol{\sigma}})\hat{\beta}_{\boldsymbol{k}}
\nonumber \\
&&
+{\displaystyle\sum_{{\boldsymbol{k}}{\boldsymbol{k}'}{\boldsymbol{q}} \sigma\lambda\beta\nu} }U^{\sigma\lambda\beta\nu}_{{\boldsymbol{k}}{\boldsymbol{k}'}{\boldsymbol{q}}}\hat{b}^{\dag}_{{\boldsymbol{k}}\sigma}\hat{b}^{\dag}_{{\boldsymbol{q}}-{\boldsymbol{k}} \lambda}\hat{b}_{{\boldsymbol{q}}-{\boldsymbol{k}'}\beta}\hat{b}_{{\boldsymbol{k}'}\nu},
\end{eqnarray}
where $\hat{\beta}^{\dag}_{\boldsymbol{k}}=(\hat{b}^{\dag}_{{\boldsymbol{k}}\uparrow}, \hat{b}^{\dag}_{{\boldsymbol{k}}\downarrow})$, $\hat{b}^{\dag}_{{\boldsymbol{k}}\sigma}$ creates a particle with Bloch momentum ${\boldsymbol{k}}$ and spin $\sigma$, and the parameters $\{\delta_{\rm eff}$, $\boldsymbol{d}_{\rm eff}$, $U^{\sigma\lambda\beta\nu}_{{\boldsymbol{k}}{\boldsymbol{k}'}{\boldsymbol{q}}}\}$ can be systematically fitted with data from full-band calculations~\cite{suppl}. Importantly, while $\boldsymbol{d}_{\rm eff}$ contains an effective in-plane Zeeman field that explicitly breaks the inversion symmetry, the effective interaction involves interesting spin-spin correlations. As shown in Figs.~\ref{fig:pd},~\ref{fig:bs}, and~\ref{fig:topo}, the effective model works reasonably well in the experimentally relevant regime with small to moderate $M_0$. We stress that the formalism discussed here is quite general and can be applied to describe high-band effects in systems with arbitrary Raman-assisted lattice SOCs.

\emph{Summary}.--
We study the behavior of a BEC under the recently realized 2D lattice SOC. While high-band effects drastically modify system properties and lead to high-band-induced ``roton'' gaps and topological phase transitions, we derive an effective two-band model, which is able to capture the high-band physics to the leading order. Our results not only offer valuable insights into the novel two-dimensional lattice spin-orbit coupling, but also provide a systematic formalism to model high-band effects in lattice systems with Raman-assisted SOC.

\emph{Acknowledgments}.--
We thank Shuai Chen, Xiong-Jun Liu for helpful discussions. This work is supported by NFRP (2011CB921200, 2011CBA00200), NKBRP (2013CB922000), NSFC (60921091, 11274009, 11374283, 11434011, 11522436, 11522545), and the Research Funds of Renmin University of China (10XNL016, 16XNLQ03).
W. Y. acknowledges support from the ``Strategic Priority Research Program(B)'' of the Chinese Academy of Sciences, Grant No. XDB01030200.

%%%%%%%%%% Merge with supplemental materials %%%%%%%%%%
\pagebreak
\widetext
\begin{center}
\textbf{\large Supplemental Materials}
\end{center}
%%%%%%%%%% Merge with supplemental materials %%%%%%%%%%
%%%%%%%%%% Prefix a "S" to all equations, figures, tables and reset the counter %%%%%%%%%%
\setcounter{equation}{0}
\setcounter{figure}{0}
\setcounter{table}{0}
\setcounter{page}{1}
\makeatletter
\renewcommand{\theequation}{S\arabic{equation}}
\renewcommand{\thefigure}{S\arabic{figure}}
\renewcommand{\bibnumfmt}[1]{[S#1]}
\renewcommand{\citenumfont}[1]{S#1}
%%%%%%%%%% Prefix a "S" to all equations, figures, tables and reset the counter %%%%%%%%%%

In this supplemental material, we provide more details on the calculation of the Bogoliubov spectrum, as well as the derivation of the effective two-band model.

\section{Bogoliubov spectrum}

Following the standard Bogoliubov theory, we expand the field operator as $\hat{\psi}_{\sigma}=\phi_{\sigma} +\delta\hat{\psi}_{\sigma}$, with the fluctuation $\delta\hat{\psi}_{\sigma}$ above the ground-state wave function $\phi_{\sigma}$. Under the basis $\{\delta\hat{\psi}_{\uparrow}, \delta\hat{\psi}_{\downarrow}, \delta\hat{\psi}_{\uparrow}^{\dag}, \delta\hat{\psi}_{\downarrow}^{\dag}\}^{T}$, the Bogoliubov Hamiltonian for the fluctuations is written as
\begin{equation}
\hat{H}_{B}=\left(\begin{array}{cc}
\hat{h}_{0}+\Gamma_{1}-\mu & \Gamma_{2}\\
\Gamma_{2}^{\ast} & \hat{h}_{0}^{\ast}+\Gamma_{1}^{\ast}-\mu
\end{array}\right),
\label{eqn:Hbogo}
\end{equation}
 where
\begin{equation}
\Gamma_{1}=\left(\begin{array}{cc}
2g\left|\phi_{\uparrow}\right|^{2}+g_{\uparrow\downarrow}\left|\phi_{\downarrow}\right|^{2} & g_{\uparrow\downarrow}\phi_{\uparrow}\phi_{\downarrow}^{\ast}\\
g_{\downarrow\uparrow}\phi_{\downarrow}\phi_{\uparrow}^{\ast} & g_{\downarrow\uparrow}\left|\phi_{\uparrow}\right|^{2}+2g\left|\phi_{\downarrow}\right|^{2}
\end{array}\right),
 \label{eqn:Gamma1}
 \end{equation}
 and
 \begin{equation}
 \Gamma_{2}=\left(\begin{array}{cc}
g\phi_{\uparrow}^{2} & g_{\uparrow\downarrow}\phi_{\uparrow}\phi_{\downarrow}\\
g_{\downarrow\uparrow}\phi_{\downarrow}\phi_{\uparrow} & g\phi_{\downarrow}^{2}
\end{array}\right).
 \label{eqn:Gamma2}
 \end{equation}
The quasi-particle spectrum can be calculated by diagonalizing the Bogoliubov Hamiltonian above using paraunitary transformations, which is equivalent to diagonalizing the following matrix
\begin{equation}
\hat{H}'_{B}=\left(\begin{array}{cc}
\hat{h}_{0}+\Gamma_{1}-\mu & \Gamma_{2}\\
-\Gamma_{2}^{\ast} & -(\hat{h}_{0}^{\ast}+\Gamma_{1}^{\ast}-\mu)
\end{array}\right).
\label{eqn:Hbogopara}
\end{equation}
The Bogoliubov spectra corresponds to the positive eigenvalues of Eq.~(\ref{eqn:Hbogopara}).

\section{The effective two-band model}

To get the effective two-band model, we add high-band related correction terms to the simple two-band Hamiltonian with both on-site and nearest-neighbor interactions. To determine the form of these correction terms, we consider a minimal multi-band model with only $s$- and $p$-orbital of the lattice potential, and perform adiabatic elimination to get a two-band model with only $s$-orbital contributions. Due to the high-band coupling, additional terms like an in-plane Zeeman field and new spin non-conserving interaction terms emerge in the model following the adiabatic elimination. We assume that these new terms should formally capture the leading-order high-band effects, and then fit their corresponding parameters using a handful of data points from the single-particle spectrum and the mean-field man-body phase diagram under the full-band calculations. As we show in the main text, the resulting effective two-band model works reasonably well in the experimentally relevant regime, when $M_0$ is not too large. In the following, we describe in detail the procedures we take in each step.

\subsection{Minimal multi-band model}

To determine the form of the correction terms. we first consider a minimal multi-band model with $s$- and $p$-orbital of the lattice potential. When we take the spin components into account, this gives rise to a six-band model. The model should capture the leading-order high-band effects with weak Raman potentials. Under the basis
$\hat{\beta}_{\boldsymbol{k}}^{\dagger}=\left(\begin{array}{cccccc}
\hat{b}_{0\boldsymbol{k}\uparrow}^{\dagger} & \hat{b}_{0\boldsymbol{k}\downarrow}^{\dagger} & \hat{b}_{1\boldsymbol{k}\uparrow}^{\dagger} & \hat{b}_{1\boldsymbol{k}\downarrow}^{\dagger} & \hat{b}_{2\boldsymbol{k}\uparrow}^{\dagger} & \hat{b}_{2\boldsymbol{k}\downarrow}^{\dagger}\end{array}\right)$,
the single-particle part of the model is given by $\sum_{\boldsymbol{k}}\hat{\beta}_{\boldsymbol{k}}^{\dagger}\hat{h}\left(\boldsymbol{k}\right)\hat{\beta}_{\boldsymbol{k}}$, where
\begin{equation}\label{sixbandH}
\hat{h}\left(\boldsymbol{k}\right)=\left(\begin{array}{ccc}
\hat{h}_{00}\left(\boldsymbol{k}\right) & -t_{\rm so,on}^{\left(0,1\right)}\hat{\sigma}_{x} & -t_{\rm so,on}^{\left(0,1\right)}\hat{\sigma}_{y}\\
-t_{\rm so,on}^{\left(0,1\right)}\hat{\sigma}_{x} & \hat{h}_{11}\left(\boldsymbol{k}\right) & t_{\rm so,on}^{\left(1,2\right)}\left(\hat{\sigma}_{y}-\hat{\sigma}_{x}\right)\\
-t_{\rm so,on}^{\left(0,1\right)}\hat{\sigma}_{y} & t_{\rm so,on}^{\left(1,2\right)}\left(\hat{\sigma}_{y}-\hat{\sigma}_{x}\right) & \hat{h}_{22}\left(\boldsymbol{k}\right)
\end{array}\right),
\end{equation}
and
\begin{equation}\label{intrabandH}
\begin{split}
&\hat{h}_{00}\left(\boldsymbol{k}\right)=\left[m_{z}-2t_{s}^{\left(0,0\right)}\cos\left(k_{x}a\right)-2t_{s}^{\left(0,0\right)}\cos\left(k_{y}a\right)\right]\hat{\sigma}_{z}-2t_{\rm so}^{\left(0,0\right)}\left[\hat{\sigma}_{y}\sin\left(k_{x}a\right)-\hat{\sigma}_{x}\sin\left(k_{y}a\right)\right],\\
&\hat{h}_{11}\left(\boldsymbol{k}\right)=\epsilon_{1}+\left[m_{z}+2t_{s}^{\left(1,1\right)}\cos\left(k_{x}a\right)\right]\hat{\sigma}_{z}+2t_{\rm so}^{\left(1,1\right)}\hat{\sigma}_{y}\sin\left(k_{x}a\right),\\
&\hat{h}_{22}\left(\boldsymbol{k}\right)=\epsilon_{1}+\left[m_{z}+2t_{s}^{\left(1,1\right)}\cos\left(k_{y}a\right)\right]\hat{\sigma}_{z}-2t_{\rm so}^{\left(1,1\right)}\hat{\sigma}_{x}\sin\left(k_{y}a\right).
\end{split}
\end{equation}
Here, the band indices $0,1$ and $2$ represent the $s$, $p_{x}$ and $p_{y}$ orbitals of the lattice potential, respectively, and $\epsilon_{1}$ is the self-energy of the $p$ orbital. Note that $\{t_{s},t_{\rm so}\}$ in the main text corresponds to $\{t_{s}^{(0,0)},t_{\rm so}^{(0,0)}\}$ above. $\hat{b}_{n\boldsymbol{k}\sigma}$ ($\hat{b}_{n\boldsymbol{k}\sigma}^{\dagger}$) is the creation (annihilation) operator of fermions with spin $\sigma$ on the $n^{\rm th}$-band with Bloch momentum $\boldsymbol{k}$. The coupling coefficients can be calculated as
\begin{equation}\label{couplingcoef}
\begin{split}
&t_{s}^{(n,n)} =\left\vert \int d\boldsymbol{r}\phi_{n\boldsymbol{j}}\left(\boldsymbol{r}\right)\left[\frac{\hbar^{2}\hat{\boldsymbol{k}}^{2}}{2m}+V\left(\boldsymbol{r}\right)\right]\phi_{n\boldsymbol{j}+\boldsymbol{e}_x}\left(\boldsymbol{r}\right)\right\vert, \\
&t_{\rm so}^{(n,n)} =\left\vert \int d\boldsymbol{r}\phi_{n\boldsymbol{j}}\left(\boldsymbol{r}\right)M\left(\boldsymbol{r}\right)\phi_{n\boldsymbol{j}+\boldsymbol{e}_x}\left(\boldsymbol{r}\right)\right\vert,\\
&t_{\rm so,on}^{(n,n{'})}=\left\vert\int d\boldsymbol{r}\phi_{n\boldsymbol{j}}\left(\boldsymbol{r}\right)M\left(\boldsymbol{r}\right)\phi_{n{'}\boldsymbol{j}}\left(\boldsymbol{r}\right)\right\vert,
\end{split}
\end{equation}
where $\phi_{n\boldsymbol{j}}\left(\boldsymbol{r}\right)$ is the $n^{\rm th}$-band Wannier function at site $\boldsymbol{j}=j_x\boldsymbol{e}_x+j_x\boldsymbol{e}_y$ with $j_{x,y}=1,2,3\ldots N_{x,y}$. Here, $N_{x,y}$ denotes the number of lattice points along the $x$- and $y$-direction, respectively, and $\boldsymbol{e}_{x,y}$ are the corresponding unit vectors.
%We also denote the self-energy of the $p$ orbitals as $\epsilon_{1}=\epsilon_{2}$, and the lattice constant as $a$.
Similar to the gauge transformation $\hat{\psi}_{\downarrow}\to e^{-ik_{0}\left(x+y\right)}\hat{\psi}_{\downarrow}$, the gauge transformation $\hat{b}_{n\boldsymbol{j}\downarrow}\to \left(-1\right)^{j_{x}+j_{y}}\hat{b}_{n\boldsymbol{j}\downarrow}$ is taken before the Hamiltonian is transformed into the Bloch-momentum space.  Note that $\hat{b}_{n\boldsymbol{j}\sigma}$ ($\hat{b}_{n\boldsymbol{j}\sigma}^{\dagger}$) is the corresponding creation (annihilation) operator of spin $\sigma$ at site $\boldsymbol{j}$. For a typical lattice depth of $V_0=4E_r$, $t_{s}^{(0,0)}=0.0855E_r$ and $t_{\rm so}^{(0,0)}/M_0=0.0545$.

In the Bloch-momentum space, the interaction part of the six-band model is given by
\begin{equation}\label{SixBInt}
V_{\rm int}=\sum_{mn}\sum_{\sigma\sigma{'}}\sum_{\boldsymbol{k}\boldsymbol{k}{'}\boldsymbol{q}}U_{mn}^{\sigma\sigma{'}}\hat{D}_{mn,\boldsymbol{k}\boldsymbol{k}{'}\boldsymbol{q}}^{\sigma\sigma{'}},
\end{equation}
where $U_{mn}^{\sigma\sigma{'}}=(2N_{x}N_{y})^{-1}g_{\sigma\sigma{'}}\Gamma_{mn}$ with $\Gamma_{mn}=(1-2\delta_{mn}/3)\int d\boldsymbol{r}\phi_{m\boldsymbol{j}}\phi_{m\boldsymbol{j}}\phi_{n\boldsymbol{j}}\phi_{n\boldsymbol{j}}$ and
\begin{equation}\label{Doperator}
\hat{D}_{mn,\boldsymbol{k}\boldsymbol{k}{'}\boldsymbol{q}}^{\sigma\sigma{'}}=\hat{b}_{m\boldsymbol{k}\sigma}^{\dagger}\hat{b}_{m\boldsymbol{q}-\boldsymbol{k}\sigma{'}}^{\dagger}\hat{b}_{n\boldsymbol{q}-\boldsymbol{k}{'}\sigma{'}}\hat{b}_{n\boldsymbol{k}{'}\sigma}+\hat{b}_{m\boldsymbol{k}\sigma}^{\dagger}\hat{b}_{n\boldsymbol{q}-\boldsymbol{k}\sigma{'}}^{\dagger}\hat{b}_{m\boldsymbol{q}-\boldsymbol{k}{'}\sigma{'}}\hat{b}_{n\boldsymbol{k}{'}\sigma}+\hat{b}_{m\boldsymbol{k}\sigma}^{\dagger}\hat{b}_{n\boldsymbol{q}-\boldsymbol{k}\sigma{'}}^{\dagger}\hat{b}_{n\boldsymbol{q}-\boldsymbol{k}{'}\sigma{'}}\hat{b}_{m\boldsymbol{k}{'}\sigma}.
\end{equation}

\subsection{Adiabatic elimination}

Starting from the six-band model above, we now adiabatically eliminate the $p$-orbital contributions to derive a two-band model. We expect that the $p$-orbital induced terms in the resulting two-band model formally capture the leading-order high-band effects.

For the single-particle case, the equation of motion (EoM) of the field operators, in the case of $g_{\sigma\sigma{'}}=0$, is given by
\begin{equation}\label{EoM}
i\hbar\frac{\partial}{\partial t}\left(\begin{array}{c}
\hat{\beta}_{0\boldsymbol{k}}\\
\hat{\beta}_{e\boldsymbol{k}}
\end{array}\right)=\left(\begin{array}{cc}
\hat{h}_{00}\left(\boldsymbol{k}\right) & B\\
B^{\dagger} & C
\end{array}\right)\left(\begin{array}{c}
\hat{\beta}_{0\boldsymbol{k}}\\
\hat{\beta}_{e\boldsymbol{k}}
\end{array}\right),
\end{equation}
where $\hat{\beta}_{0\boldsymbol{k}}^{\dagger}=\left(\begin{array}{cc}
\hat{b}_{0\boldsymbol{k}\uparrow}^{\dagger} & \hat{b}_{0\boldsymbol{k}\downarrow}^{\dagger}\end{array}\right)$, $\hat{\beta}_{e\boldsymbol{k}}^{\dagger}=\left(\begin{array}{cccc}
\hat{b}_{1\boldsymbol{k}\uparrow}^{\dagger} & \hat{b}_{1\boldsymbol{k}\downarrow}^{\dagger} & \hat{b}_{2\boldsymbol{k}\uparrow}^{\dagger} & \hat{b}_{2\boldsymbol{k}\downarrow}^{\dagger}\end{array}\right)$, $B=\left(-\begin{array}{cc}t_{\rm so,on}^{\left(0,1\right)}\hat{\sigma}_{x} & -t_{\rm so,on}^{\left(0,1\right)}\hat{\sigma}_{y}\end{array}\right)$ and
\begin{equation}\label{BCmatrix}
C=\left(\begin{array}{cc}
\hat{h}_{11}\left(\boldsymbol{k}\right) & t_{\rm so,on}^{\left(1,2\right)}\left(\hat{\sigma}_{y}-\hat{\sigma}_{x}\right)\\
t_{\rm so,on}^{\left(1,2\right)}\left(\hat{\sigma}_{y}-\hat{\sigma}_{x}\right) & \hat{h}_{22}\left(\boldsymbol{k}\right)
\end{array}\right).
\end{equation}

In the limit of large $V_0$ and small $M_0$, the bands in the $p$-orbital can be adiabatically eliminated. We may then require $\partial\hat{\beta}_{e\boldsymbol{k}}/\partial t\approx0$, and derive an effective two-band Hamiltonian for the six-band model by substituting $\hat{\beta}_{e\boldsymbol{k}}\approx-C^{-1}B^{\dagger}\hat{\beta}_{0\boldsymbol{k}}$ into the EoM of $\hat{\beta}_{0\boldsymbol{k}}$: $\hat{h}_{00}^{\rm eff}=\hat{h}_{00}-BC^{-1}B^{\dagger}$. Furthermore, we can write $C=C_1+C_2$, with $C_{1}=\epsilon_{1}$. The Taylor expansion of $C^{-1}$ with respect to $C_{2}$ up to the first-order term gives $C^{-1}=\left(C_{1}+C_{2}\right)^{-1}\approx C_{1}^{-1}-C_{1}^{-2}C_{2}$. Considering the case with small $m_z$, $C^{-1}$ approximately equals to the Taylor expansion of $C^{-1}$ with respect to $C_{2}$ up to the first-order term $C_{1}^{-1}-C_{1}^{-2}C_{2}$
\begin{equation}\label{TaylorE}
C^{-1}\approx\frac{1}{\epsilon_{1}}-\frac{1}{\epsilon_{1}^{2}}\left(\begin{array}{cc}
\hat{h}_{11}\left(\boldsymbol{k}\right) & t_{\rm so,on}^{\left(1,2\right)}\left(\hat{\sigma}_{y}-\hat{\sigma}_{x}\right)\\
t_{\rm so,on}^{\left(1,2\right)}\left(\hat{\sigma}_{y}-\hat{\sigma}_{x}\right) & \hat{h}_{22}\left(\boldsymbol{k}\right)
\end{array}\right).
\end{equation}

We can then obtain an effective two-band single-particle Hamiltonian for the six-band model, which is only exact in the limit of large $V_0$, and small $m_z$ and $M_0$
\begin{align}
\hat{h}_{00}^{\rm eff}=\delta^{\rm eff}_{00}+\boldsymbol{d}_{00}^{\rm eff}\cdot\boldsymbol{\hat{\sigma}},
\end{align}
where $\delta_{00}^{\rm eff}=-2\chi\epsilon_{1}$ and
\begin{equation}\label{CoEM}
\begin{split}
&d_{00,x,y}^{{\rm eff}}=\pm 2 \left[m_{\perp}^{(0,0)}+\left(t_{\rm so}^{\left(0,0\right)}+\delta t_{\rm so}^{\left(0,0\right)}\right)\sin\left(k_{y,x}a\right)\right],\\
&d_{00,z}^{{\rm eff}}=\left(m_{z}+\delta m_{z}^{(0,0)}\right)-2\left(t_{s}^{\left(0,0\right)}+\delta t_{s}^{\left(0,0\right)}\right)\left[\cos\left(k_{x}a\right)+\cos\left(k_{y}a\right)\right],
\end{split}
\end{equation}
with $\chi=t_{\rm so,on}^{\left(0,1\right)2}/\epsilon_{1}^{2}$, $m_{\perp}^{(0,0)}=\chi t_{\rm so,on}^{\left(1,2\right)}$, $\delta t_{\rm so}^{\left(0,0\right)}=\chi t_{\rm so}^{\left(1,1\right)}$, $\delta m_z^{(0,0)}=-2\chi m_z$ and $\delta t_{s}^{\left(0,0\right)}=\chi t_{s}^{\left(1,1\right)}$.
By comparing $\hat{h}_{00}^{\rm eff}$ and $\hat{h}_{00}$, we see that apart from corrections to the existing terms in the simple two-band model,
an additional high-band induced in-plane Zeeman field $m_{\perp}^{(0,0)}(\boldsymbol {e}_x-\boldsymbol{e}_y)$ emerges in the effective model.

We now proceed to discuss the form of the interaction corrections. When the interaction coefficients $g_{\sigma\sigma{'}}$ are finite but not too large, we can neglect the interaction terms in the EoM of $\hat{\beta}_{e\boldsymbol{k}}$ and directly write down the effective interaction Hamiltonian by using the relationship $\hat{\beta}_{e\boldsymbol{k}}\approx-C^{-1}B^{\dagger}\hat{\beta}_{0\boldsymbol{k}}$,
\begin{equation}\label{EffInt}
V_{{\rm int},00}^{\rm eff}\approx\sum_{\boldsymbol{k}\boldsymbol{k}{'}\boldsymbol{q},\sigma\beta\lambda\nu}\mathcal{U}_{\boldsymbol{k}\boldsymbol{k}{'}\boldsymbol{q}}^{\sigma\beta\lambda\nu}\hat{b}_{0\boldsymbol{k}\sigma}^{\dagger}\hat{b}_{0\boldsymbol{q}-\boldsymbol{k}\beta}^{\dagger}\hat{b}_{0\boldsymbol{q}-\boldsymbol{k}{'}\lambda}\hat{b}_{0\boldsymbol{k}{'}\nu},
\end{equation}
where
\begin{equation}\label{EffIntCoef}
\begin{split}
\mathcal{U}_{\boldsymbol{k}\boldsymbol{k}{'}\boldsymbol{q}}^{\sigma\beta\lambda\nu}=&\sum_{mn,\rho\rho{'}}U_{mn}^{\rho\rho{'}}(\Sigma_{\boldsymbol{k},m\rho,\sigma}^{\ast}\Sigma_{\boldsymbol{q}-\boldsymbol{k},m\rho{'},\beta}^{\ast}\Sigma_{\boldsymbol{q}-\boldsymbol{k}{'},n\rho{'},\lambda}\Sigma_{\boldsymbol{k}{'},n\rho,\nu}+\Sigma_{\boldsymbol{k},m\rho,\sigma}^{\ast}\Sigma_{\boldsymbol{q}-\boldsymbol{k},n\rho{'},\beta}^{\ast}\Sigma_{\boldsymbol{q}-\boldsymbol{k}{'},m\rho{'},\lambda}\Sigma_{\boldsymbol{k}{'},n\rho,\nu}\\
&+\Sigma_{\boldsymbol{k},m\rho,\sigma}^{\ast}\Sigma_{\boldsymbol{q}-\boldsymbol{k},n\rho{'},\beta}^{\ast}\Sigma_{\boldsymbol{q}-\boldsymbol{k}{'},n\rho{'},\lambda}\Sigma_{\boldsymbol{k}{'},m\rho,\nu})
\end{split}
\end{equation}
and $\Sigma_{\boldsymbol{k}}^{\dagger}=\left(\begin{array}{cc} \hat{\sigma}_{0} & \Sigma_{e\boldsymbol{k}}^{\dagger}\end{array}\right)$ with
\begin{equation}\label{SigMatr}
\Sigma_{e\boldsymbol{k}}=\frac{t_{\rm so,on}^{\left(0,1\right)}}{\epsilon_{1}^{2}}\left(\begin{array}{c}
-t_{\rm so,on}^{\left(1,2\right)}\hat{\sigma}_{0}+\epsilon_{1}\hat{\sigma}_{x}-i\left[m_{z}+2t_{s}^{\left(1,1\right)}\cos\left(k_{x}a\right)\right]\hat{\sigma}_{y}+i\left[t_{\rm so,on}^{\left(1,2\right)}+2t_{\rm so}^{\left(1,1\right)}\sin\left(k_{x}a\right)\right]\hat{\sigma}_{z}\\
t_{\rm so,on}^{\left(1,2\right)}\hat{\sigma}_{0}+\epsilon_{1}\hat{\sigma}_{y}+i\left[m_{z}+2t_{s}^{\left(1,1\right)}\cos\left(k_{y}a\right)\right]\hat{\sigma}_{x}+i\left[t_{\rm so,on}^{\left(1,2\right)}+2t_{\rm so}^{\left(1,1\right)}\sin\left(k_{y}a\right)\right]\hat{\sigma}_{z}
\end{array}\right).
\end{equation}
We want to emphasize the coefficients of every interaction terms all can be analytically written down. Physically, high-band corrections lead to complicated spin non-conserving interactions in the effective model.

\subsection{Effective two-band model}

%In the previous subsection, we derive a simple effective two-band model for the minimum multi-band model in the weak-coupling limit: large $V_0$, and small $M_0$ and $m_z$. However, the situation becomes very complicated when more higher bands are considered.
%To proceed, we assume that all the high-band corrections on the $s$ orbital can be packaged into the corrections from a set of bands similar to the $p$ bands. This assumption is plausible since all the higher bands have similar hopping and interaction terms. Therefore the effective two-band Hamiltonian for the full-band model is assumed to take the same form as that for the minimum multi-band model (Eq.~(\ref{CoEM}) and (\ref{EffInt})), with a set of coefficients that are determined by high-band effects. By fixing these coefficients using results from full-band calculations under finite $M_0$ and $m_z$, we expect that the effective model is applicable in a larger parameter regime. Particularly, the effective two-band model should still captures the leading-order high-band corrections so long as the energy gap between the $p$ bands and the $s$ bands remains large.

\begin{figure}[tbp]
\begin{center}
\includegraphics[width=12cm]{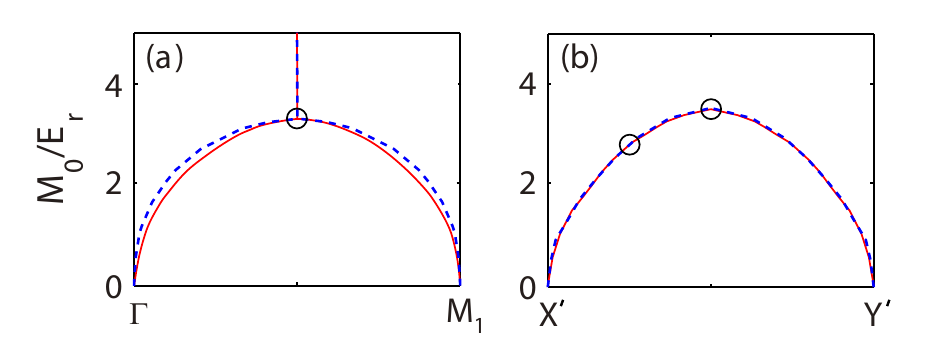}
\caption{(Color online) {Comparison of the lowest-energy points (a) and the Dirac points (b) between the full-band model (red solid curves) and the effective two-band model (blue dashed curves). The open circles denote the points which are selected as the fitting points.}}
\end{center}
\label{fig:extremum}
\end{figure}

With the form of the effective two-band model fixed, we now determine the coefficients of the high-band correction terms by fitting with results from full-band calculations. These coefficients include five single-body parameters $\epsilon_1$, $t_s^{(1,1)}$, $t_{\rm so}^{(1,1)}$, $t_{\rm so,on}^{(0,1)}$ and $t_{\rm so,on}^{(1,2)}$, and one interaction parameter $\Gamma_{01}$. Note that as the coefficients $\Gamma_{mn}$ ($m,n=1,2$) are associated with fourth-order terms of the high-band operators, they are of higher-order smallness. We therefore do not fit these coefficients, but rather use their original values, $\Gamma_{11}=\Gamma_{22}\approx0.0458E_r$,$\Gamma_{12}=\Gamma_{21}\approx0.0409E_r$, together with the fitting parameter $\Gamma_{01}^{\prime}$ in calculating the interaction coefficients in the final effective model. The five single-body parameters are fixed by fitting the effective model at four characteristic points marked by the open circles in Fig. 1 and Fig. 4(a) in the main text, and by fitting the energy shift of the Dirac points of the single-particle spectrum with the same parameters as the marked point in Fig. 1(a). The fitting gives a set of linear equations which are easily solved to obtain the single-body parameters. $\Gamma_{01}$ is fixed with the characteristic point which is marked with the red open circle in Fig. 2 in the main text. The values of the original and the fitted high-band-involved parameters are listed in Tab.~\ref{table1}.

\begin{table}
\begin{center}
\begin{tabular}{|l|l|l|l|l|l|l|}
\hline
original         & $t_{\rm so,on}^{(0,1)}$  & $\epsilon_1$ & $t_s^{(1,1)}$ & $t_{\rm so}^{(1,1)}$ & $t_{\rm so,on}^{(1,2)}$ & $\Gamma_{01}$\\
\hline
$*/E_r$          & 0.397$M_0/E_r$ & 2.97 & 0.465 & 0.161$M_0/E_r$ & 0.230$M_0/E_r$ & 0.0961\\
\hline
fitted            & $t_{\rm so,on}^{(0,1)'}$  & $\epsilon_1^{\prime}$ & $t_s^{(1,1)'}$ & $t_{\rm so}^{(1,1)'}$ & $t_{\rm so,on}^{(1,2)'}$  & $\Gamma_{01}^{\prime}$\\
\hline
$*/E_r$          & 0.405$M_0/E_r$ & 2.45 & 0.582 & 0.0887$M_0/E_r$ &  0.243$M_0/E_r$ & 0.0557\\
\hline
\end{tabular}
\end{center}
\caption{The values of the original and fitted parameters in the effective two-band model.}
\label{table1}
\end{table}

Typically, the tight-binding approximation with on-site interaction terms is used to derive the two-band model. Although the strength of the nearest-neighbour interaction terms are far smaller than that of the on-site interaction terms, the average energy contribution of these nearest-neighbour interaction terms is considerable for typical lattice depths in current experiments. Indeed, the ground-state phase diagram under a simple two-band model apparently differ from that of the full-band model at $M_0=0$. In order to quantitatively reproduce the results of the full-band model, these nearest-neighbour interaction terms need to be included for the $s$ bands in the effective two-band model. We want to emphasize that all conclusions regarding the high-band effects in this paper are not affected by this correction. By making the substitution using the fitted parameters
\begin{equation}\label{ParamReplac}
\left\{ \epsilon_{1},t_{s}^{(1,1)},t_{\rm so}^{(1,1)},t_{\rm so,on}^{(0,1)},t_{\rm so,on}^{(1,2)},\Gamma_{01}\right\} \longrightarrow\left\{ \epsilon_{1}^{\prime},t_{s}^{(1,1)'},t_{\rm so}^{(1,1)'},t_{\rm so,on}^{(0,1)'},t_{\rm so,on}^{(1,2)'},\Gamma_{01}^{\prime}\right\},
\end{equation}
and
\begin{equation}\label{NearNeigh}
\begin{split}
&U_{00}^{\rho\rho'}\longrightarrow U_{00,\boldsymbol{k}\boldsymbol{k}{'}\boldsymbol{q}}^{\prime \rho\rho{'}}=U_{00}^{\rho\rho{'}}+\\
&+2\eta U_{00}^{\rho\rho{'}}\sum_{l=x,y}\left\{ \xi_{\rho}\cos\left(k_{l}a\right)+\xi_{\rho{'}}\cos\left[\left(q_{l}-k_{l}\right)a\right]+\xi_{\rho{'}}\cos\left[\left(q_{l}-k_{l}^{\prime}\right)a\right]+\xi_{\rho}\cos\left(k_{l}^\prime{}a\right)\right\}
\end{split}
\end{equation}
in Eqs. (\ref{CoEM}) and (\ref{EffIntCoef}) respectively, where $\xi_{\uparrow\downarrow}=\pm1$ and $\eta=\Gamma_{00}^{\prime}/\Gamma_{00}\approx-0.0152$ with $\Gamma_{00}^{\prime}$ the overlap integral of one $s$-orbital Wannier function at site $\boldsymbol{j}$ and three other $s$-orbital Wannier functions at site $\boldsymbol{j}+\boldsymbol{e}_{x}$, we obtain the final effective two-band model in the main text,
\begin{equation}\label{eq:EffH0}
\hat{H}_{\rm eff}=\sum_{\boldsymbol{k}}\hat{\beta}_{0\boldsymbol{k}}^{\dagger}\left(\delta_{\rm eff}+\boldsymbol{d}_{\rm eff}\cdot\boldsymbol{\hat{\sigma}}\right)\hat{\beta}_{0\boldsymbol{k}}+\sum_{\boldsymbol{k}\boldsymbol{k}{'}\boldsymbol{q},\sigma\beta\lambda\nu}U_{\boldsymbol{k}\boldsymbol{k}{'}\boldsymbol{q}}^{\sigma\beta\lambda\nu}\hat{b}_{0\boldsymbol{k}\sigma}^{\dagger}\hat{b}_{0\boldsymbol{q}-\boldsymbol{k}\beta}^{\dagger}\hat{b}_{0\boldsymbol{q}-\boldsymbol{k}{'}\lambda}\hat{b}_{0\boldsymbol{k}{'}\nu}.
\end{equation}
Note that for simplicity, we write $\{\hat{\beta}_{0\boldsymbol{k}},\hat{b}_{0\boldsymbol{k}\sigma}\}$ as $\{\hat{\beta}_{\boldsymbol{k}},\hat{b}_{\boldsymbol{k}\sigma}\}$ in the main text. Here, the energy shift $\delta_{\rm eff}$ and the vector $\boldsymbol{d}_{\rm eff}$ are given by $\delta_{\rm eff}=-2\chi^{\prime}\epsilon_{1}^{\prime}$ and
\begin{equation}\label{dvector}
\begin{split}
&d_{{\rm eff},x,y}=\pm 2 \left[m_{\perp}+\left(t_{\rm so}+\delta t_{\rm so}\right)\sin\left(k_{y,x}a\right)\right],\\
&d_{{\rm eff},z}=\left(m_{z}+\delta m_{z}\right)-2\left(t_{s}+\delta t_{s}\right)\left[\cos\left(k_{x}a\right)+\cos\left(k_{y}a\right)\right],
\end{split}
\end{equation}
with $\chi^{\prime}=(t_{\rm so,on}^{\left(0,1\right)'}/\epsilon_{1}^{\prime})^2$, $m_{\perp}=\chi^{\prime}t_{\rm so,on}^{\left(1,2\right)'}$, $\delta t_{\rm so}=\chi^{\prime}t_{\rm so}^{\left(1,1\right)'}$, $\delta m_z=-2\chi^{\prime}m_z$ and $\delta t_s=\chi^{\prime}t_{s}^{\left(1,1\right)'}$. The upper index $(0,0)$ is dropped here to be consistent with the expressions in the main text. For the lattice depth $V_0=4E_r$, $m_{\perp}\approx0.0066M_0^3/E_r^2$, $\delta t_{\rm so}\approx0.0024 M_0^3/E_r^2$, $\delta m_z\approx-0.0547M_0^2 m_z/E_r^2$, $\delta t_s\approx0.0159M_0^2/E_r$ and $\delta_{\rm eff}\approx-0.1338M_0^2/E_r$.

The effective interaction coefficient $U_{\boldsymbol{k}\boldsymbol{k}{'}\boldsymbol{q}}^{\sigma\beta\lambda\nu}$ is given by
\begin{equation}\label{Redifined_EffIntCoef}
\begin{split}
&U_{\boldsymbol{k}\boldsymbol{k}^{'}\boldsymbol{q}}^{\sigma\beta\lambda\nu}=3U_{00,\boldsymbol{k}\boldsymbol{k}^{'}\boldsymbol{q}}^{'\sigma\beta}\delta_{\sigma\nu}\delta_{\beta\lambda}+\sum_{mn\neq00,\rho\rho^{'}}U_{mn}^{'\rho\rho^{'}}(\Sigma_{\boldsymbol{k},m\rho,\sigma}^{'\ast}\Sigma_{\boldsymbol{q}-\boldsymbol{k},m\rho^{'},\beta}^{'\ast}\Sigma_{\boldsymbol{q}-\boldsymbol{k}^{'},n\rho^{'},\lambda}^{'}\Sigma_{\boldsymbol{k}^{'},n\rho,\nu}^{'}\\
&+\Sigma_{\boldsymbol{k},m\rho,\sigma}^{'\ast}\Sigma_{\boldsymbol{q}-\boldsymbol{k},n\rho^{'},\beta}^{'\ast}\Sigma_{\boldsymbol{q}-\boldsymbol{k}^{'},m\rho^{'},\lambda}^{'}\Sigma_{\boldsymbol{k}^{'},n\rho,\nu}^{'}+\Sigma_{\boldsymbol{k},m\rho,\sigma}^{'\ast}\Sigma_{\boldsymbol{q}-\boldsymbol{k},n\rho^{'},\beta}^{'\ast}\Sigma_{\boldsymbol{q}-\boldsymbol{k}^{'},n\rho^{'},\lambda}^{'}\Sigma_{\boldsymbol{k}^{'},m\rho,\nu}^{'}),
\end{split}
\end{equation}
where $U_{mn}^{\prime \rho\rho{'}}=(2N_{x}N_{y})^{-1}g_{\sigma\sigma{'}}\Gamma_{mn}^{\prime}$ with $\Gamma_{mn}^{\prime}=\Gamma_{mn}$ as $m,n\neq0$, and $\Sigma_{\boldsymbol{k}}^{'\dagger}=\left(\begin{array}{cc} \hat{\sigma}_{0} & \Sigma_{e\boldsymbol{k}}^{'\dagger}\end{array}\right)$ with
\begin{equation}\label{Redifined_Sigma}
\Sigma_{e\boldsymbol{k}}^{\prime}=\frac{t_{\rm so,on}^{\left(0,1\right)'}}{(\epsilon_{1}^{\prime})^2}\left(\begin{array}{c}
-t_{\rm so,on}^{\left(1,2\right)'}\hat{\sigma}_{0}+\epsilon_{1}{'}\hat{\sigma}_{x}-i\left[m_{z}+2t_{s}^{\left(1,1\right)'}\cos\left(k_{x}a\right)\right]\hat{\sigma}_{y}+i\left[t_{\rm so,on}^{\left(1,2\right)'}+2t_{\rm so}^{\left(1,1\right)'}\sin\left(k_{x}a\right)\right]\hat{\sigma}_{z}\\
t_{\rm so,on}^{\left(1,2\right)'}\hat{\sigma}_{0}+\epsilon_{1}{'}\hat{\sigma}_{y}+i\left[m_{z}+2t_{s}^{\left(1,1\right)'}\cos\left(k_{y}a\right)\right]\hat{\sigma}_{x}+i\left[t_{\rm so,on}^{\left(1,2\right)'}+2t_{\rm so}^{\left(1,1\right)'}\sin\left(k_{y}a\right)\right]\hat{\sigma}_{z}
\end{array}\right),
\end{equation}
and the identity matrix $\hat{\sigma}_0$.


\begin{thebibliography}{99}
\bibitem{footnotenew} Under a two-band model, the locations of the Dirac points are immune to changes of $M_0$. The locations of the lowest-energy states are essentially immune to changes of $M_0$, except for a transition at $M_0/E_r\approx 2.2$. For $M_0/E_r<2.2$, the lowest-energy points are at the $\Gamma$ and ${\rm M}_{i=1,2,3,4}$ points; for $M_0/E_r>2.2$, the lowest-energy points are at the mid-point on the $\Gamma$--${\rm M}_i$ line, i.e., with a four-fold degeneracy.
\bibitem{footnote} For the full-band calculation here, we have included the lowest twenty two bands. Our results have already converged under this high-band cutoff.
\bibitem{footnote2} The fine structure in the Chern number phase diagram of the Bogoliubov spectrum is also present, albeit with slightly different boundaries, in a simple two-band model. Thus this feature is intrinsic of the 2D SOC rather than high-band induced.
\end{thebibliography}
\end{document}